\documentclass[aps,prl,showpacs,groupedaddress,twocolumn]{revtex4}
\usepackage{amsmath,amssymb}
\usepackage{graphicx} 
\usepackage{dcolumn} 
\usepackage{bm} 

\begin{document}
\draft
\title{Electric control of spin in monolayer WSe$_2$ field effect transistors}
\author{Kui Gong$^{1,2\dag}$, Lei Zhang$^{1\ddag}$, Dongping Liu$^1$, Lei Liu$^3$, Yu Zhu$^3$, Yonghong Zhao$^{1,4}$ and Hong Guo$^1$}
\address{$^1$Department of physics, McGill University, Montreal, H3A 2T8, Canada\\
$^2$School of Materials Science and Engineering, University of Science and Technology Beijing,
Beijing, 100083 China\\
$^3$Nanoacademic Technologies Inc., Brossard, QC, J4Z 1A7, Canada\\
$^4$College of Physics and Electronic Engineering, Sichuan Normal University, Chengdu, 610068 China}


\begin{abstract}

We report a first principles theoretical investigation of quantum transport in monolayer WSe$_2$ field effect transistor (FET). Due to a strong spin-orbit interaction (SOI) and the atomic structure of the two-dimensional (2D) lattice, monolayer WSe$_2$ has an interesting electronic structure that exhibits Zeeman-like up-down spin texture near the $K$ and $K'$ points of the Brillouin zone. In a FET, the gate electric field induces an extra, externally tunable SOI that re-orients the spins into a Rashba-like texture thereby realizing electric control of the spin. Quantum transport is modulated by the spin texture, namely by if the spin orientation of the carrier after the gated channel region, matches or miss-matches that of the FET drain electrode. The carrier current in the FET is labelled both the spin index and the valley index, realizing spintronics and valleytronics in the same device.

\end{abstract}

\maketitle

Electronic materials in reduced dimension have attracted great attention for decades. The newest member of such material is the two dimensional (2D) transition-metal dichalcogenides (TMDC). Since the exfoliation of \emph{monolayer} TMDC (ML-TMDC) three years ago\cite{Splendiani,Mak1}, very  interesting electronic and optical properties of these materials have been already discovered both experimentally and
theoretically\cite{MoS-transistor,Xiao,Zeng,Mak2,Cao,Yuan}.  TMDC is in the form of MX$_2$ where M denotes heavy elements such as Mo, W, and X denotes S, Se, etc.. The most important properties of several ML-TMDC, for instance WSe$_2$ and MoS$_2$, are the direct band gap in the visible frequency range\cite{Splendiani,Mak1} and the strong spin-orbit interaction (SOI). ML-TMDC materials have honeycomb lattice shown in Fig. 1(a) and in the momentum space there are two inequivalent valleys at $K$ and $K'$ of the first Brillouin zone (BZ, Fig. 1(a)). Due to the well-separation of $K$ from $K'$, it was proposed\cite{Xiao,Zeng,Mak2,Cao} that the valley index $\tau=K,K'$ may be used as quantum numbers for \emph{valleytronics}. At the same time, the strong SOI are fundamentally important for \emph{spintronics}. Being able to realize both valleytronics and spintronics in a single material is a very exciting new opportunity and ML-TMDC may well be the emerging electronic material for new generations of nanoelectronics. It is the purpose of this work to theoretically investigate fundamental properties of quantum transport in ML-TMDC material WSe$_2$ in the form of a field effect transistor (FET).

\begin{figure}
\centering
\includegraphics[width=8.5cm,height=5.5cm]{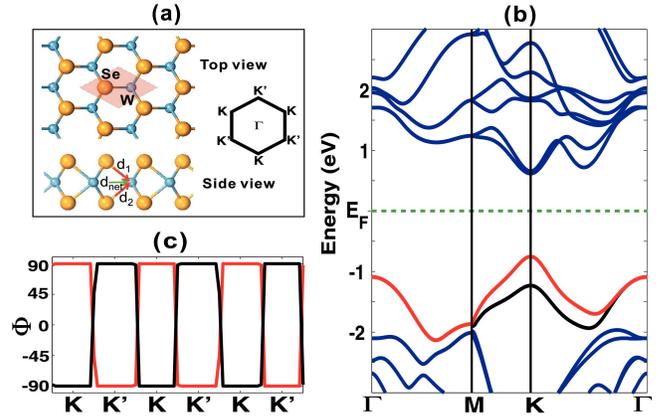}
\caption{(Color online)(a). Top view and side view of monolayer WSe$_2$ and its corresponding first BZ with inequivalent $K$ and $K'$ points. In-plane dipoles are found around the W atoms as indicated by the red arrows labelled $\textbf{d}_1,\textbf{d}_2$ which add up to a net dipole $\textbf{d}_{net}$. (b). The calculated electronic band structure of the monolayer WSe$_2$. The Fermi level is indicated by the horizontal dashed green line. (c). Tilting angle $\Phi$ along the edge of the first Brillouin zone for the uppermost two spin splitting valence bands, where the red and dark lines correspond to the two valence bands in panel (b) marked in the same colors, respectively.} \label{fig1}
\end{figure}

In particular, by atomistic first principles analysis we found that the spins in WSe$_2$ can be well controlled by an external \emph{electric} field - here by a gate voltage of the FET, and such a control has direct consequences to quantum transport properties of the device. Achieving efficient electric control of spin is a long-sought goal of spintronics \cite{Sahoo,Nitta,Matsuyama,Grundler}, for WSe$_2$ it is due to not only the strong SOI but also to its 2D lattice structure. Conceptually, SOI is proportional to $\hat{{\mathbf s}}\cdot(\nabla V({\bf r})\times {\mathbf k})$, where $\hat{{\mathbf s}}$ is the spin and ${\mathbf k}$ the momentum of the carrier while $ {\bf E}=-\nabla V({\bf r})$ the electric field seen by the carrier. One may view $(\nabla V({\bf r})\times {\mathbf k})\equiv \textbf{B}_{eff}$ as an \emph{effective} magnetic field acting on spin $\hat{{\mathbf s}}$. For monolayer WSe$_2$, ${\mathbf k}$ is in the 2D plane; $\textbf{E}$ is dominated by an internal electric dipole field\cite{Yuan} (see Fig. 1(a)) thus also in the plane, giving rise to a $\textbf{B}_{eff}$ largely perpendicular to the plane to orient the spins into a Zeeman like up-down texture\cite{Yuan} in most regions of the BZ. For WSe$_2$ FET, gate voltage adds an additional electric field $\textbf{E}_{ext}$ which is perpendicular to the 2D plane:  $(\textbf{E}_{ext} \times {\mathbf k})$ is thus oriented inside the plane which attempts to orient the spins into a Rashba like texture. The spin orientations can thus be controlled by $\textbf{E}_{ext}$. As a result, after a carrier traverses the gated channel of the FET, its spin orientation is modulated by the gate voltage and its transmission probability is large or small depending on the spin orientation being matched or mismatched to the spin orientations in the drain contact. Both equilibrium and nonequilibrium quantum transport properties are therefore influenced. For the WSe$_2$ FET, the current is not only labelled by the spin index $s$ but also the valley index $\tau$: $I=I_{s,\tau}$, both spin-current and valley-current are possible outcomes of this interesting device.

Before analyzing FET, we briefly discuss important basic properties of monolayer WSe$_2$. To this end we relaxed the atomic structure of the material by density functional theory (DFT) with the projector augmented plane wave (PAW) method\cite{Blochl} as implemented in the VASP package\cite{vasp,footnote1}. Fig. 1(b) plots the calculated band structure showing a direct gap\cite{footnote1a} and a lift of spin degeneracy at the top of the valence band and bottom of  conduction band along the M-K and K-$\Gamma$ lines. Due to the absence of inversion symmetry of monolayer WSe$_2$, the spin splitting reaches a maximum value $\sim 480$meV at the top of valence band (black and red lines in Fig. 1(b)) in $K$ ($K'$) points. Such a splitting is very substantial and much larger than other ML-TMDC materials.

From the calculated charge $Q_{nk}$ of the Bloch state for band $n$ and wave vector $k$, we determine a spin polarization vector $\mathbf{P}=(P_x,P_y,P_z)$ by decomposing $Q_{nk}$ using the Pauli matrix $\sigma$\cite{Zhao}, $Q_{nk}=Q\mathbf{I} + P_x\sigma_x+P_y\sigma_y+P_z\sigma_z$ where $\mathbf{I}$ is the unit matrix. From $\textbf{P}$ we obtain the titling angle $\Phi = arctan[P_z/(P_x^2+P_y^2)^{1/2}]$ which characterizes the spin texture: $\Phi=\pm 90^o$ means the spins are Zeeman-like, i.e. perpendicular to the 2D plane. Fig. 1(c) shows $\Phi$ for the top two valance bands along the edge of the first BZ. At the $K$ or $K'$ point, spin polarization is clearly Zeeman-like\cite{Yuan}: one band has spin up ($\Phi = 90^o$) and the other spin down ($-90^o$). Spin polarization of the same band alternates between $\pm 90^o$ going from $K$ to $K'$ to $K$ etc. (see Fig. 1(a)) having a three-fold rotational symmetry which reflects  the $D^1_{3h}$ crystal symmetry of monolayer WSe$_2$. We note in passing that going away from the $K,K'$ valleys, $\Phi$ reduces from $\pm 90^o$ which means the spins tilt away from the perpendicular direction. A calculated ``map" of $\Phi$ in the BZ is included in the Supplemental Materials\cite{EPAPS}.
\begin{figure}
\centering
\includegraphics[width=8.0cm,height=7.4cm,clip=true]{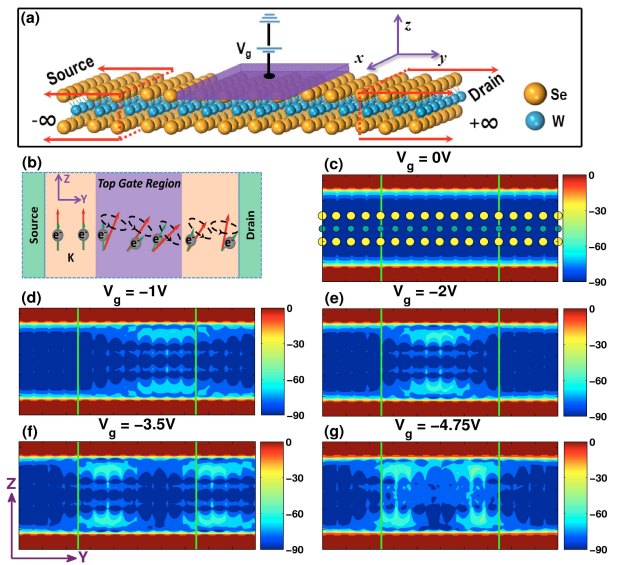}
\caption{(Color online) (a). Structure of the monolayer WSe$_2$ FET with a top gate. The dotted lines indicate source and drain which extend to $y=\pm\infty$. The W and Se atoms are shown in blue and yellow. Periodic boundary condition is assumed in the $x$ direction. (b). Schematic plot of a carrier incoming from the $K$ valley and traversing the gated channel region where its spin is modulated by the gate voltage. (c) - (g): Titling angle $\Phi({\bf r})$ of the scattering state $\psi(E,k)$ in the $y-z$ plane averaged over the $x$-direction, for several gate voltages. The side bars give the color coding for the value of $\Phi({\bf r})$. In (c), W and Se atoms are indicated as blue and yellow balls. The two vertical green lines mark the gated region.} \label{fig2}
\end{figure}

Having understood the spin texture in monolayer WSe$_2$, we now analyze the electric control of the texture in FET shown in Fig. 2(a). The FET is a two-probe open structure of WSe$_2$ where the channel region is controlled by gate voltage $V_g$, the source/drain extend to $y=\pm \infty$ where bias voltage is applied and current $I_{\tau,s}$ collected. For open device structures under nonequilibrium, the state-of-the-art first principles method is to carry out DFT within the Keldysh nonequilibrium Green's function (NEGF) formalism\cite{jeremy1}. Here, the WSe$_2$ FET requires self-consistent NEGF-DFT analysis to include SOI and non-collinear spin, together with self-consistent determination of the bias and gate potentials. Our calculation is by the first principles quantum transport package Nanodcal\cite{jeremy1,derek1,nanodcal2}. For technical details we refer interested readers to Refs.\cite{jeremy1,derek1} and computation details to Ref.\cite{footnote2}. Again, we analyze the spin texture using polarization ${\mathbf P}$, but for FET it is calculated from scattering states which are eigenstates of the open device Hamiltonian. Thus ${\mathbf P}(E,k,{\mathbf r})=(\langle S_x\rangle,\langle S_y\rangle,\langle S_z\rangle)$ in real space where $\langle S_{\alpha}({\mathbf r})\rangle=\langle\psi(E,k,{\mathbf r})|s_{\alpha}|\psi(E,k,{\mathbf r})\rangle$ $(\alpha=x,y,z)$ is the expectation value of spin operators at position ${\mathbf r}$,  $\psi(E,k,{\mathbf r})$ is the two-component scattering spinor wave function which we obtain from the NEGF-DFT calculation\cite{jeremy1}, and $s_{\alpha}$ is the corresponding Pauli matrix.

As presented above, the Zeeman type spin splitting occurs in the valence band (Fig. 1(b)), hence we focus on quantum transport in the energy range $[-1.75,-0.75]$eV. Fig. 2(b) plots a qualitative spin evolution where, as discussed above, due to $V_g$ an extra and in-plane SOI is induced, thus the incoming Zeeman-like spins from the source are rotated toward the 2D plane when traversing the gated region, resulting to a Rashba-like spin texture. Note we only plotted the up-spins in Fig. 2(b): due to time reversal symmetry exactly the same happens to the down-spins. Very importantly, the gated region of the FET acts as a large ``defect" along the pristine WSe$_2$ to break the translational symmetry, thereby opening up the phase space for spin rotation by $V_g$.

Quantitatively, the electric control of ${\mathbf P}$ is vividly shown in Fig. 2(c-g) which are obtained for transport along the zigzag direction of the WSe$_2$, at energy $E=-0.77$eV and near the $K'$ valley. At $V_g=0$, Fig. 2(c) shows that the Zeeman-like spin texture ($\Phi(y,z)=-90^{\circ}$, blue regions in the plot) is preserved in the entire FET\cite{footnote4}. Increasing $V_g$, polarization becomes $|\Phi({\mathbf r})| < 90^{\circ}$ (the lighter blue regions), namely the spins swing toward the 2D x-y plane: a consequence of competition between the $V_g$-induced Rashba and the intrinsic Zeeman spin-splitting. Importantly, the rotation of the spin texture depends on the value of $V_g$. At $V_g=-1$ and $-3.5$V (Fig. 2(d,f)), $\Phi({\mathbf r})$ is substantially away from $90^o$ after exiting the gated region; on the other hand for $V_g=-2$ and $-4.75$V (Fig. 3(e,g)), $\Phi({\mathbf r})$ is close to $90^o$ after the gated region. We emphasize that while spins are rotated by $V_g$, the time reversal symmetry is not broken since up- and down-spins are rotated exactly opposite ways. The gate control of spin texture has significant consequences to quantum transport as we show below.

Having established the electric control of spin texture in the FET, we now calculate both equilibrium and nonequilibrium quantum transport properties of the 2D transistor. We consider two different FETs where transport is along the armchair or zigzag directions of the WSe$_2$ lattice and mediated by the valence bands. Fig. 3(a) plots the calculated zero bias conductance $G$ versus electron energy $E$ for various gate voltages\cite{footnote5}. $G$ reduces as $E$ increases toward the band gap of the material, because of reduction of the number of incoming channels at larger $E$, consistent with the band structure (Fig. 1(b)). For our pure WSe$_2$ FET and at a given $E$, $V_g$ monotonically reduces $G$ because it changes the potential of the channel relative to that of the source/drain, in the energy range up to the top of valence band at  $-1.1$eV (black curve in Fig. 1(b)). Very interesting behavior occurs in the valance band range $E=[-0.9, -0.75]$eV (inset of Fig. 3(a)) where incoming electrons are located in the $K$ or $K'$ valleys whose spin polarization is Zeeman-like. In this energy range, $G$ is no longer monotonic in $V_g$ but is modulated in an oscillatory manner by it.

The modulation of $G$ by $V_g$ is clearly shown in Fig. 3(b). The modulation period depends on FET being zigzag or armchair, but also depends on the length of the gated region. We found that the modulation period is doubled when the length of the gated region is reduced to half. Furthermore, fixing the length of the gate, $G$ modulates with the length of the central region of the FET which contains the gated region plus the buffer layers on either side of the gated region (see Fig. 2(a)). These results - happening in the valence band where large SOI spin splitting occurs, clearly indicate that they are related to the spin degree of freedom. Indeed, for the $E=-0.77$eV curve in Fig. 3(b), the maxima of $G$ occur at $V_g=-2$ and $-4.75$V, corresponding to the spin textures in Fig. 2(e,g) where spin rotation is such that $\Phi({\mathbf r})$ is close to $90^o$ after the gated region. On the other hand, the minima of $G$ occur at $V_g=-1$V and $-3.5$V, corresponding to the spin textures in Fig. 2(d,f) where $\Phi({\mathbf r})$ is substantially away from $90^o$ after the gated region. Because spins in the drain of the FET are Zeeman-like, carriers reaching there with $\Phi \sim 90^o$ can easily transmit to the outside world, resulting to a large $G$. The opposite is true for carriers having $\Phi \neq 90^o$. We conclude that due to electric control of the spin by SOI, conductance of the WSe$_2$ FETs can be modulated by a gate voltage of several volts. Varying the length of the gated region, the necessary $V_g$ for the modulation can also be changed. In this regard, the WSe$_2$ FET possesses the main device feature of the long-sought Datta-Das spin-transistor.\cite{Datta,Mireles}

\begin{figure}
\centering
\includegraphics[width=8.5cm,height=6cm]{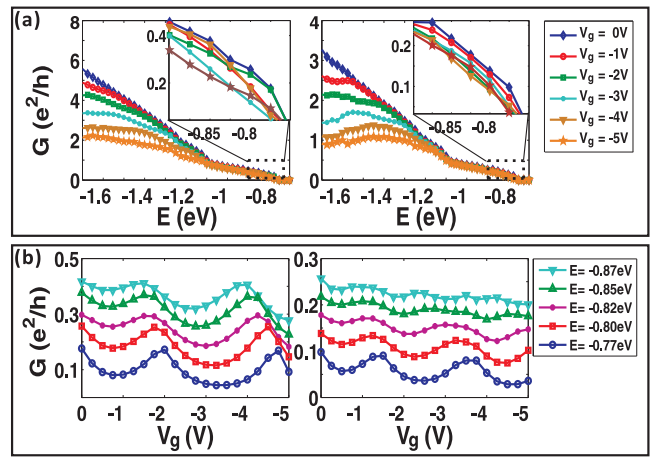}
\caption{(Color online) (a). Conductance $G$ versus energy $E$ for the zigzag FET (left) and armchair FET (right) at several gate voltages. Inset:  $G$ in the zoom-in energy range $[-0.9,-0.75]$eV. (b). $G$ versus the gate voltage $V_g$ for the zigzag FET (left) and armchair FET (right) at several energy values in the valence range.} \label{fig3}
\end{figure}

So far the results were obtained from parameter-free first principles methods, in the following we construct an analytical model to qualitatively explain the observed conductance modulations in the FET. To this end we start from a two-band effective Hamiltonian proposed in Ref.\cite{Xiao} for the $K,K'$ valleys of the WSe$_2$, $H_{2B} = at(\tau k_x \hat{\sigma}_x + k_y \hat{\sigma}_y) + (\Delta/2)\hat{\sigma}_z - \lambda \tau \hat{s}_z(\hat{\sigma}_z-1)/2,$
where $\tau=\pm1$ is the valley index, $a$ is the lattice constant, $t$ the hopping integral, $\Delta$ the energy gap, $2\lambda$ is the spin splitting at the top valence band, Pauli matrix $\hat{\sigma}_{x,y,z}$ is a pseudospin indexing the $A=d_{z^2}$ and $B=(d_{x^2-y^2}\pm id_{xy})/\sqrt{2}$ orbitals and $\hat{s}_{x,y,z}$ is the real spin. $(k_x,k_y)$ is the wave vector measuring from the $K$ or $K'$ points of the BZ. For our FET the gate voltage induces an extra Rashba-like SOI and breaks the translational symmetry, as a result intervalley scattering between $K$ and $K'$ becomes possible. Therefore, we augment $H_{2B}$ by a Rashba SOI term $H_{Rashba}$ which, in the basis of $A, B$ orbitals, has been worked out before\cite{lu} and presented in the Supplemental Material\cite{EPAPS}. Therefore, our model Hamiltonian becomes $H=H_{2B}+H_{Rashba}$. From our \emph{ab initio} results we estimated the Rashba strength and found it to be $\sim 10^{-11}$eVm which is not large. We can treat $H_{Rashba}$ as a perturbation to $H_{2B}$ and solve the problem analytically\cite{EPAPS}. In particular, for a spin-up carrier at the $K$ valley with $k_x=0$ injected from the source into the gated region of length $L$, we solve the wave function emerging from the gated region from which the probability of this carrier ending up into the spin-up or -down channel is obtained. As shown in the Supplemental Material\cite{EPAPS}, this probability is found to be proportional to cosine or sine functions of a variable that depends on $L$, $V_g$ and SOI parameters. The effective model $(H_{2B}+H_{Rashba})$ thus reproduces the spin modulation and confirm the importance of the induced $H_{Rashba}$ term. In a real FET, incoming carriers with momentum ${\bf k}$ of the $K$ valley can be scattered into $\bf k'$ of the $K'$ valley in the gated region making the transport more complicated so that the conductance modulation is not simply a sinusoidal function (see Fig. 3(b)).

\begin{figure}
\centering
\includegraphics[width=8.5cm,height=3.5cm]{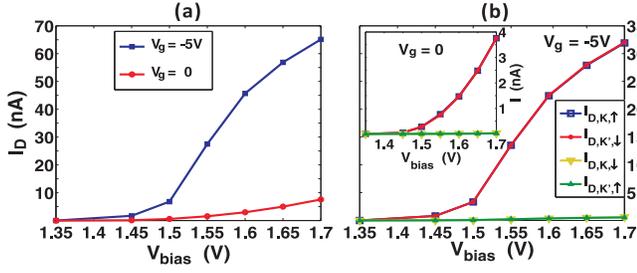}
\caption{(Color online) (a) Drain current $I_{D}$ versus bias voltage with and without a gate voltage. (b). Different components of the drain current $I_{D,\tau,s}$ versus bias voltage with and without a gate voltage.} \label{fig4}
\end{figure}

We found that the WSe$_2$ FET has fascinating non-equilibrium transport properties when an external bias is applied. In particular the presence of valley ($\tau$) and spin ($s$) degrees of freedom makes the current $I$ a four-index quantity in terms of valley-spin degrees of freedom,
\begin{equation}
I_{\alpha,\tau,s}=\frac{e^2}{h} \int^{\mu_S}_{\mu_D} dE \sum_{{\bf k}\in \tau} T(E,{\bf k},s) \label{current1}
\end{equation}
where $\alpha$ labels source and drain (S,D) of the FET, $\mu_{S,D}$ the chemical potentials of them. In our numerical calculations, the external bias is applied in the drain, the applied bias window is such that valence band with only $K,\uparrow$ and $K',\downarrow$, conduction band with $\tau=K,K'$, $s=\uparrow,\downarrow$ will participate transport. Therefore, $I_{\alpha}$ can be expressed as:
\begin{equation}
\begin{split}
I_{S}&=I_{S,K,\uparrow} + I_{S,K',\downarrow},\\
I_{D}&=I_{D,K,\uparrow} + I_{D,K,\downarrow} + I_{D,K',\uparrow} + I_{D,K',\downarrow}.
\end{split}
\end{equation}
Note that the source current $I_{S}$ is composed of two components $I_{S,K,\uparrow}=I_{S,K',\downarrow}=I_{S}/2$, because the spin and valley indexes are locked together in the source of the FET. After the gated region, due to scattering the drain current has four components.

Figure \ref{fig4} plots the calculated drain current for a zigzag FET, armchair FET gives very similar results. Fig. \ref{fig4}(a) shows the total drain current $I_{D}$ versus bias voltage $V_b$. Since the monolayer WSe$_2$ is a semiconductor having an intrinsic band gap, $I_{D}$ is very small when $V_g=0$ (red curve) and it is greatly enhanced when a finite $V_g$ is applied (blue curve), showing the typical transistor character. Figure \ref{fig4}(b) plots different components of the drain current $I_{D,\tau,s}$. Even though the current is mainly contributed by $I_{D,K,\uparrow}$ and $I_{D,K',\downarrow}$, the other two components still account for about 1.7\% of the current.

In the drain region of the FET, by summing up the quantum numbers one can collect a valley-current $I_{D,\tau}=I_{D,\tau,\uparrow}+I_{D,\tau,\downarrow}$, and a spin-current $I_{D,s}=I_{D,K,s}+I_{D,K',s}$. For our FET and due to time reversal symmetry, the total valley-current and total spin-current vanish: $(I_{D,K}-I_{D,K'})/2=0$, $(I_{D,\uparrow}-I_{D,\downarrow})/2=0$. Nevertheless, if the time reversal symmetry is broken, e.g. by a magnetic field or magnetic impurity, a non-vanishing total valley- and spin-current can be produced in the FET which will be an extremely interesting outcome.

In summary, by atomistic first principles calculations we have shown that 2D-TMDC WSe$_2$ is a very interesting electronic material for realizing a special FET that has both spintronics and valleytronics characteristics. Due to the strong SOI and the atomic structure of  the 2D WSe$_2$ lattice, monolayer WSe$_2$ exhibits Zeeman-like up-down spin texture near the $K$ and $K'$ points of the Brillouin zone. In a FET, the gate electric field induces an extra, externally tunable SOI that re-orients the spins into a Rashba-like texture, realizing electric control of the spin. Quantum transport is modulated by the spin texture, namely depending on whether or not the carrier spin orientation after the gated region matches that of the FET drain electrode. A gate voltage of a few volts is adequate to tune the spins and affect the carrier current. A simple effective Hamiltonian is proposed to account for the qualitative behavior of the spin modulation.

{\bf Acknowledgments.} We thank Dr. Ferdows Zahid for bringing our attention to TMDC, and Dr. Wang Yao for discussions of the valley physics. We gratefully acknowledge financial support by NSERC of Canada and and University Grant Council (AoE/P-04/08) of the Government of HKSAR (H.G.), and the China Scholarship Council (K.G.). We thank CLUMEQ and CalcuQuebec for providing computation facilities.

\noindent
$^\dag$: kui.gong@mail.mcgill.ca\\
$^\ddag$: zhanglei@physics.mcgill.ca

\vspace{2cm}

\noindent {\bf {\underline{EPAPS: Supplemental Material}} for
``Electric field control of spin in monolayer WSe$_2$ transistors", by Kui Gong et al.}

\vspace{0.5cm}
In this Supplemental Material, we present theoretical details of two issues of the main text. Section I provides calculated results of the tilting angle $\Phi$ at the top valence band in the first Brillouin zone (BZ)  surrounding the $\Gamma$ point of the monolayer WSe$_2$. Section II gives more details of solving the analytical model $H_{2B}+H_{Rashba}$ to qualitatively explain the observed conductance modulations in the FET.

\section{I. Titling angle in the Brillouin zone }

As explained in the main text, due to the intrinsic local electric field of the in-plane dipoles in the monolayer WSe$_2$ lattice (see Fig.1a and also Ref.[8] of the main text), spins are organized in the  Zeeman-like texture: one band has spin up ($\Phi = 90^o$) and the other spin down ($-90^o$) at the $K$ or $K'$ point of the BZ.

We have also calculated $\Phi$ in the first BZ surrounding the $\Gamma$ point as shown in Fig.1 of this EPAPS document. Indeed, large areas of the BZ have $\Phi \sim \pm 90^o$ (red or blue regions) especially away from the $\Gamma$ point ($k_x=k_y=0$). Near the $\Gamma$ point,  $\Phi$ is less than $|\pm 90^o|$ which means the spins tilt away from the perpendicular direction of the 2D plane. The angle $\Phi$ preserves three fold rotation symmetry in the BZ which reflects the lattice symmetry of the monolayer WSe$_2$.

\section{II. The two band model}

In this section we propose an effective two-band model Hamiltonian to account for the conductance modulation found in the first principles calculations. To this end we start from the two-band effective Hamiltonian proposed in Ref.\cite{Xiao} for the $K,K'$ valleys of the TMDC material,
\begin{equation}
H_{2B} = at(\tau k_x \hat{\sigma}_x + k_y \hat{\sigma}_y) + \frac{\Delta}{2}\hat{\sigma}_z - \lambda \tau \hat{s}_z\frac{\hat{\sigma}_z-1}{2},\label{eqH}
\end{equation}
where $\tau=\pm1$ denotes $K$ or $K'$ valley, respectively; $a$ is the lattice constant; $t$ is the hopping integral; $\Delta$ is the energy gap; $2\lambda$ is the spin splitting at the top valence band; the Pauli matrix $\hat{\sigma}_{x,y,z}$ is the pseudospin, indexing the $A=d_{z^2}$ and $B=(d_{x^2-y^2}\pm id_{xy})/\sqrt{2}$ orbitals; and $\hat{s}_{x,y,z}$ is for the real spin. Note that $(k_x,k_y)$ is the wave vector measuring from $K$ or $K'$ point.

\begin{figure}
\centering
\includegraphics[width=5.5cm,height=4.5cm]{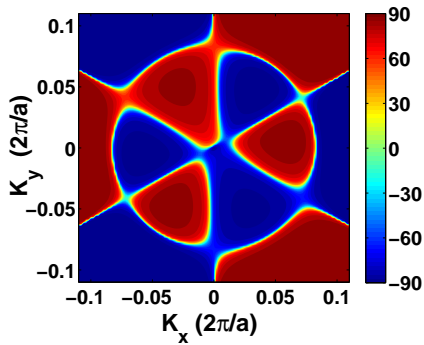}
\caption{(Color online) Tilting angle $\Phi$ of the top valence band in the first BZ surrounding the $\Gamma$ point ($k_x=k_y=0$). The side-bar is the color coding for the values of $\Phi$. Large areas of the BZ have $\Phi \sim \pm 90^o$ (red or blue regions). Near the $\Gamma$ point  $\Phi$ is less than $|\pm 90^o|$ which means the spins tilt away from the perpendicular direction of the 2D plane.}
\label{fig-supp}
\end{figure}

We shall be interested in the conductance modulation behavior occurring in the energy range of the valance band, $E=[-0.9, -0.75]$eV where incoming electrons are located at  the top valence band $K$ or $K'$ valleys whose spin polarization is Zeeman-like. Solving the eigen-value equation $H_{2B}|v,\tau,s\rangle=E^0_{v}|v,\tau,s\rangle$, one can obtain the eigenvalues of the uppermost valence band ($v$),
\begin{equation}
E^0_{v}=\frac{\lambda}{2}-\sqrt{\frac{(\Delta-\lambda)^2}{4}+(a\times t\times k)^2},
\end{equation}
where $k=\sqrt{k_x^2+k_y^2}$. The corresponding eigenvectors (of the uppermost valence band) have the form
\begin{equation}
\begin{split}
|v,K,+\rangle&=\left(\begin{array}{c}
        1\\
        0\\
\end{array}\right)\otimes
\left(\begin{array}{c}
        \sin(\frac{\theta}{2})\\
        -\cos(\frac{\theta}{2})e^{i\phi_{k}}\\
\end{array}\right),\\
|v,K',-\rangle&=\left(\begin{array}{c}
        0\\
        1\\
\end{array}\right)\otimes
\left(\begin{array}{c}
        \sin(\frac{\theta}{2})\\
        \cos(\frac{\theta}{2})e^{-i\phi_{k}}\\
\end{array}\right),\label{eqVector}
\end{split}
\end{equation}
where $\phi_{k}=\arctan (\frac{k_y}{k_x})$ and $\theta=\arccos(\frac{\Delta-\lambda}{\sqrt{(\Delta-\lambda)^2+(2 a t k)^2}})$. Note that two states $|v,K,+\rangle$ and $|v,K',-\rangle$ are degenerate with the same eigen-energy $E^0_{v}$.

For our FET, as discussed in main text the gate voltage induces an extra Rashba-like SOI and breaks the translational symmetry, as a result intervalley scattering between $K$ and $K'$ in the gate region becomes possible. From results of our \emph{ab initio} calculation, we found that for an incoming state ${\bf k}$ the deviation of angle $\Phi$ from $90^o$ becomes smaller when the momentum difference $\delta {\bf k}={\bf k}'-{\bf k}$ becomes larger (where ${\bf k}\in K, {\bf k'}\in K'$), this means scattering preferentially occurs between $K$ and $K'$ valleys at the same energy. The Rashba SOI induced by the gate voltage is expressed as
\begin{equation}
H_{Rashba}=\frac{\hbar}{4m_e^2c^2}\hat{{\mathbf s}}\cdot(\nabla V({\bf r})\times {\mathbf p}),\label{eqH1}
\end{equation}
where $V({\bf r})$ is the electric field due to the gate voltage. Treating scattering in the 2D plane and take Fourier transform of Eq. (\ref{eqH1}) from real space to ${\bf k}$ space and, in the basis of $A$ and $B$ orbitals,\cite{lu}
\begin{equation}
H_{Rashba}({\bf k},{\bf k'})=\left[\begin{array}{cccc}
         0 & 0 & U^A_{-} & 0\\
         0 & 0 & 0 & U^B_{-}\\
         U^A_{+} & 0 & 0 & 0\\
         0 & U^B_{+} & 0 & 0\\
\end{array}\right],
\end{equation}
with scattering matrix elements $U_{\pm}^{A/B}=i\int d{\bf r} V^{A/B}({\bf r})e^{i({\bf k'}-{\bf k})\cdot {\bf r}}{\bf k'}\times{\bf k}\cdot (\hat{x}\pm i\hat{y})$, where we have used notation $V^{A/B}({\bf r})=\sum_{{\bf R}}V({\bf r-R})$ (${\bf R}\in A/B$) to denote the potential at position ${\bf R}$ and the coefficient $\frac{\hbar}{4m_e^2c^2}$ in Eq. (\ref{eqH1}) is incorporated into $V({\bf r-R})$.

We propose a simple model by combining $H_{2B}$ and $H_{Rashba}$ for our FET:
\begin{equation}
H\ =\  H_{2B}+ H_{Rashba}\ .
\end{equation}
From our \emph{ab initio} results we can estimate a Rashba strength which is found to be of the order $10^{-11}$eVm. This is small and can be treated perturbatively. According to the degenerate perturbation theory, we obtain the following eigen-value equation,
\begin{equation}
\left[\langle \xi|H_{2B}+H_{Rashba}|\xi\rangle-E\right]\phi=0,
\end{equation}
where $\xi\equiv\left(\begin{array}{c}|v,K,+\rangle \\ |v,K',-\rangle \\ \end{array}\right)$. It is easy to find the eigenvalues to be $E^{\pm}=E^0_{v}\pm |AA|$ where  $AA\equiv\sin^2(\frac{\theta}{2})U^A_{-}-\cos^2(\frac{\theta}{2})U^B_{-}e^{-2i\phi_{k}}$. The corresponding eigen-function is $\phi=\left(\begin{array}{c}\alpha \\ \pm \beta \\ \end{array}\right)$ where $\alpha=\beta=\frac{1}{\sqrt{2}}$. The wave function is therefore $\psi^{\pm}=\alpha|v,K,+\rangle\pm \beta|v,K',-\rangle$ for eigenenergy $E^{\pm}$.

For simplicity, we consider that a spin-up electron from the $K$ valley with $k_x=0$ is injected from the source of a zigzag FET into the gate region (length $L$). As a result of the Rashba SOI induced by the gate voltage, the wave function emerging from the gate region can be represented by $\psi^+ e^{ik_{y1}L}+\psi^-e^{ik_{y2}L}$, where $k_{y1}$ and $k_{y2}$ satisfies the corresponding eigenenergies $E^{+}(k_{y1})=E^{-}(k_{y2})$. Then the probability of projecting it into spin-up ($|v,K,+\rangle$) or -down ($|v,K', -\rangle$) with $E^0_{v}(k_y)$ is proportional to $cos^2[(k_{y2}-k_{y1})L/2]$ or $sin^2[(k_{y2}-k_{y1})L/2]$ with different coefficients. Therefore, the total conductance is proportional to these oscillatory factors. This shows that the induced Rashba SOI gives rise to the conductance modulations shown in Fig. 3(b) of the main text. It is evident that the length of gate region ($L$) affects the period of the conductance modulation. In addition, the gate voltage affects matrix elements $U^{A/B}$ and hence the wave vector $k_{y1,y2}$ that are obtained from the eigenenergies, namely the period of the conductance modulation is also influenced by $V_g$ as we found in the \emph{ab initio} data.


\begin{thebibliography}{99}
\bibitem{footnote1a}
Using the relaxed atomic structure, we calculated the electronic structure by both VASP\cite{vasp} and Nanodcal\cite{nanodcal2}, the former is a plane wave DFT method while the latter is a LCAO DFT method. Good agreement was found and Fig.\ref{fig1}(b) is the band structure obtained by Nanodcal. Further details of Nanodcal calculations are listed in Ref.\cite{footnote2}.
\bibitem{footnote4}
In Fig. (3c-g), in the vacuum region along the $z$ direction, $\Phi(y,z)=0$ by definition since these regions are outside of the WSe$_2$ material.
\bibitem{footnote5}
The energy $E$ and momentum $k_x$ resolved transmission coefficient is obtained from the Green's functions as $T(E,k_x)=\mathrm{Tr}[\Gamma_{S,k_x}G^r_{k_x}\Gamma_{D,k_x}G^a_{k_x}]$
where $k_x$ is the wave vector sampled in the one dimensional BZ (for transport). $\Gamma_{S/D,k_x}$ is the linewidth function of the source (S) and drain (D); $G^{r,a}_{k_x}$ are the retarded/advanced Green's functions. The total transmission is $T(E)=\int dk_x T(E,k_x)$ and the conductance $G(E)=\frac{e^2}{h}T(E)$.
\end{thebibliography}
\end{document}